\documentclass{article}
\usepackage{amsfonts}
\usepackage{amssymb}
\usepackage{amsmath}
\usepackage{amsthm}
\usepackage{mathtools}
\begin{document}

\title{Non-primary square roots in massive gravity }

\author{Alexey Golovnev\\
{\small \it Centre for Theoretical Physics, The British University in Egypt,}\\
{\small \it El Sherouk City, Cairo 11837, Egypt}\\
{\small  agolovnev@yandex.ru} } 
\date{}

\maketitle

\begin{abstract}

Non-linear dRGT massive and bimetric gravities are complicated theories constructed in terms of square roots of matrices. Apart from the technical issues of successfully working with such square roots, there is also a problem of their non-uniqueness.  There are claims in the literature that one should better use the principal root. This is a very reasonable conclusion. However, the motivation they give for it is that otherwise there would be non-primary square roots violating the general covariance. In this paper, I would like to show that, if properly understood, the non-primary square roots are also perfectly covariant. At the same time, I recall the relatively old observation that the real problem with such square roots lies in perturbation theory around them. In terms of matrices, it simply does not exist. In terms of the elementary symmetric polynomials used in the Lagrangian density, it is not analytic. Moreover, the non-principal square roots are more prone to getting into the complex domain.

\end{abstract}

\section{Introduction}

Among other modified gravity models, massive and bimetric gravities have got a very peculiar place with an exciting history. One can find the old history of massive gravity in the review paper \cite{Kurt}, pretty much biased to the field theory viewpoint with the helicity 2 carrier taken above the beautiful geometry behind. The paper was written soon after the final construction of the dRGT non-linear action \cite{dRGT} and its generalisation \cite{HR1}, and approximately at the time of the first full proof of ghost-freedom \cite{HR2} as well as extensions to arbitrary reference metrics \cite{HR3} and bimetric theories \cite{HR4}. These developments, with all the background and history and many related topics, were then reviewed by one of the founders \cite{Claudia}. Another nice review, also with history and with some hints of subsequent developments, can be found in the paper \cite{Angnis}.

Massive gravity with the physical metric $g$ and a fiducial metric $f$ in $n$ spacetime dimensions of Minkowski signature $(-,+,+,\ldots,+)$ can be formulated by the following action principle:
\begin{equation}
\label{genmas}
S=\int  \left(R(g) + \sum\limits_{i=0}^n \beta_i e_i\left(\sqrt{g^{-1}f}\right)\right)\sqrt{\mathrm{-det} (g)}  d^n x
\end{equation}
where $\beta_i$ is a coupling constant and $e_i\left(\sqrt{g^{-1}f}\right)$ is the $i$-th elementary symmetric polynomial of eigenvalues of the matrix $\sqrt{g^{-1}f}$, that is such a matrix $\mathfrak R$ that ${\mathfrak R}^2 = g^{-1}f$. As always, $R(g)$ is the scalar curvature of the metric $g$. A bimetric theory with the second dynamical metric $f$ can be constructed by adding its own Einstein-Hilbert term $R(f)\sqrt{\mathrm{-det} (f)} $ to the Lagrangian density (\ref{genmas}).

Recall that, if $\lambda_1, \lambda_2, \ldots, \lambda_n$ are eigenvalues of an $n\times n$ matrix $\mathcal M$, taken with their algebraic multiplicity, then by definition
\begin{equation}
\label{defe}
e_i ({\mathcal M}) \equiv \sum\limits_{k_1 < k_2 < \cdots < k_i} \lambda_{k_1}\lambda_{k_2} \cdots \lambda_{k_i}
\end{equation}
and $e_0\equiv 1$. For example, in 3D one would have
$$e_0=1, \qquad e_1=\lambda_1 + \lambda_2 + \lambda _3, \qquad e_2=\lambda_1 \lambda_2 + \lambda_1 \lambda_3+ \lambda_2 \lambda _3, \qquad e_3 = \lambda_1 \lambda_2 \lambda_3. $$
Obviously, the values of $e_i$ do not depend on the chosen numbering of the eigenvalues, and $e_n({\mathcal M})=\mathrm{det} (\mathcal M)$ in any dimension $n$. We also define it such that $e_m({\mathcal M})\equiv 0$ for any $m>n$.

Massive gravity research was very popular for a few years after the inception of dRGT models. Then the popularity started declining. I would say, to a large extent, it was because the framework is a rather complicated one, while its benefits for phenomenology are not obvious; even though, in a full-fledged bimetric setting, viable cosmological models can be constructed \cite{cosm1, cosm2, cosm3}, while the massive cousin of the graviton might probably be used for Dark Matter \cite{Fede}.

Despite its low popularity nowadays, the technical issues of massive gravity are of interest. One of the corresponding themes is non-uniqueness of square roots. Already some time ago, the strategy of using the principal root only was motivated \cite{MiKo} by allegedly non-covariant nature of a square root with different square roots of equal eigenvalues in different Jordan blocks. These claims are also repeated in new works around bimetric gravity and its extensions \cite{ncovn}. My aim in this paper is to show that there is no violation of covariance associated with such square roots, though there are other severe problems with them.

\section{The square roots and their covariance}

We consider matrices over the field of complex numbers. In complex numbers, every non-degenerate matrix does have a square root. For the degenerate case, it is not necessarily true, with the simplest counterexample being $\left(\begin{matrix}
0 & 1\\
0 & 0
\end{matrix}\right)$. Fortunately, the matrix $g^{-1}f$ must be non-degenerate, so that  $ 0\notin \sigma\left({g^{-1}f}\right)$ -- zero is not in its spectrum. It means that all $\sigma\left({g^{-1}f}\right)\ni\lambda_i \neq 0$, and they come in a number of Jordan blocks equal to their geometric multiplicity $n_g$:
\begin{equation}
\label{normf}
{\mathcal M} = {\mathcal S} \left(J_{d_1} (\lambda_1) \bigoplus J_{d_2} (\lambda_2) \bigoplus\ \ldots \ \bigoplus J_{d_{n_g}} (\lambda_{d_{n_g}} )\right){\mathcal S}^{-1}, \qquad d_1 + d_2 + \ldots + d_{n_g} = n
\end{equation}
where the symbol $A \bigoplus B$ means a block-diagonal matrix $\mathrm{diag} (A,B)$ and the matrix $\mathcal S$ represents a similarity transformation. If all the blocks are 1-dimensional, $J_1(\lambda)=(\lambda)$, then we deal with a diagonalisable matrix. Otherwise, a Jordan block $J_d(\lambda)$ with $d>1$ is given in components as $J^i_i=\lambda$ (no summation), $J^i_{i+1}=1$ (first upper diagonal), and all other components vanishing.  A proof of the spectral theorem (\ref{normf}) can be found in any textbook on matrices, as well as sketched in the Appendix of our paper \cite{meunr}.

If the blocks are 1-dimensional, their square roots are simply $\pm(\sqrt{\lambda})$, while otherwise they are given by upper-triangular matrices
$${\sqrt{J}}^i_i =\pm\sqrt{\lambda},  \qquad{\sqrt{J}}^i_{i+k} = \frac{{\sqrt{J}}^i_i }{k! \lambda^k} \prod\limits_{j=0}^{k-1}\left(\frac12 - j\right)$$
spectrally equivalent to $\pm J(\sqrt{\lambda})$. We can then immediately adopt a finite number of square roots:
\begin{equation}
\label{theroot}
\sqrt{\mathcal M}=  {\mathcal S} \left( \bigoplus\limits_i \left(\pm \sqrt{J_{d_i}(\lambda_i)}\right) \right){\mathcal S}^{-1}.
\end{equation}
These are the obvious solutions for the square root equation, with the familiar freedom of signs. 

The classical option would be to choose the principal square root. Namely, assuming that there are no real negative eigenvalues, we represent each eigenvalue as $\lambda = r e^{i\phi}$ with $-\pi<\phi<\pi$ and define its square root as $\sqrt{\lambda} = \sqrt{r} e^{i\phi/2}$. On one hand, this is an arbitrary choice of the branch cut. On the other hand, for real-valued matrices in this case, all complex-conjugate $\lambda$-s produce complex-conjugate $\sqrt{\lambda}$-s and positive-definite operators have positive-definite square-roots.

At the same time, in cases of non-principal square roots, one might also have two equal eigenvalues with different choices of square roots for them. These particular cases of non-principal square roots were emphatically called non-primary ones in the paper \cite{MiKo}. Before discussing their reasons for dismissing them, let us first see what is special in them for the matrix analysis.

Having found the square roots above (\ref{theroot}), we might want to look for other options. Without changing the spectral decomposition, its similarity transformation to the matrix $\sqrt{\mathcal M}$ might be different:
\begin{equation}
\label{moreroot}
\sqrt{\mathcal M}=  {\mathcal T} \left( \bigoplus\limits_i \left(\pm \sqrt{J_{d_i}(\lambda_i)}\right) \right){\mathcal T}^{-1}.
\end{equation}
Equating the two potentially  different versions (\ref{theroot},\ref{moreroot}) of $\left(\sqrt{\mathcal M}\right)^2$ to each other, we get the following condition:
$$\left( \bigoplus\limits_i J_{d_i}(\lambda_i) \right){\mathcal T}^{-1} {\mathcal S} = {\mathcal T}^{-1} {\mathcal S} \left( \bigoplus\limits_i J_{d_i}(\lambda_i)\right).$$
In other words, though written in a slightly different way, the extra freedom is available when there exists a similarity transformation that changes the chosen $\sqrt{\mathcal M}$ without affecting the matrix $\mathcal M$; or put differently, there must exist a matrix ${\mathcal T}^{-1} {\mathcal S} $ that commutes with the spectral decomposition of $\mathcal M$, but not with that of $\sqrt{\mathcal M}$.

It is not difficult to see \cite{meunr} that a substantially non-trivial matrix ${\mathcal T}^{-1} {\mathcal S} $ is possible if, and only if, there exist two different Jordan blocks with equal eigenvalues. For the sake of simplicity, let us assume that the matrix $\mathcal M$ is diagonalisable and see what happens. If every eigenspace is 1-dimensional only, then all its eigenvalues are different from each other, and the only matrices commuting with it are the ones that are diagonal in the very same basis. This is something I do not call substantially non-trivial. It does not add any freedom to the choice of a square-root matrix: its eigenvectors are all the same as the initial matrix possessed, and the eigenvalues are $\pm\sqrt{\lambda_i}$.

If an eigenvalue $\lambda$ is of a higher multiplicity, then there exists a bigger subspace ${\mathbb V}_{\lambda}$ in the vector space $\mathbb V$ in which the linear operator of the matrix $\mathcal M$ acts:
\begin{equation}
\label{bei}
{\mathbb V}_{\lambda}\subset {\mathbb V}; \qquad {\mathcal M}:\quad   {\overrightarrow v} \mapsto \lambda {\overrightarrow v} \iff {\mathbb V}_{\lambda} \ni {\overrightarrow v}.
\end{equation}
This is the eigenspace of the eigenvalue $\lambda$. The linear space ${\mathbb V}_{\lambda}$ is also an invariant subspace of the square-root operator $\sqrt{\mathcal M}$. However, now we might separate it into two different eigenspaces:
\begin{equation}
\label{sei}
{\mathbb V}_{\lambda}= {\mathbb V}_{+\sqrt{\lambda}} \bigoplus {\mathbb V}_{-\sqrt{\lambda}}\ ; \qquad {\sqrt{\mathcal M}}:\quad   {\overrightarrow v} \mapsto \pm\sqrt{\lambda} {\overrightarrow v} \iff {\mathbb V}_{\pm\sqrt{\lambda}} \ni {\overrightarrow v}.
\end{equation}
The substantially non-trivial matrices ${\mathcal T}^{-1} {\mathcal S} $ are those that are not diagonal in the chosen basis of $\mathbb V_{\lambda}$. They do anyway commute with $\mathcal M$, but not necessarily so with $\sqrt{\mathcal M}$. In other words, the algebraic variety of possible square roots is about choosing different decompositions (\ref{sei}) of the big eigenspace (\ref{bei}).

The authors of the paper \cite{MiKo} claim that such square roots, dubbed non-primary by them, do violate general covariance. Their logic is as follows. If we perform a change of coordinates on the spacetime manifold, it acts as a similarity transformation of the matrix $g^{-1}f$. In the cases of all different eigenvalues, we will be forced to apply the same similarity transformation to the square root $\sqrt{g^{-1}f}$ found by the formula (\ref{theroot}). Then they simply claim that it is not the case for non-primary square roots. This is not true. Applying a similarity transformation to the representation (\ref{normf}), we would naturally do the same for the square root (\ref{theroot}), whatever eigenvalue equalities were or were not there. And this is the correct thing to do.

What is true is that some transformations do not change the shape of the matrix $\mathcal M$ while changing its non-primary square roots. Maybe, this is what had made the authors of \cite{MiKo} claim that non-primary square roots violate diffeomorphism invariance. In reality, it only means that the formula (\ref{theroot}), if taken naively, is not a good definition of a square-root matrix. For a non-primary square root, its result depends not only on the matrix $\mathcal M$ itself and the choices of $\pm\sqrt{\lambda_i}$ but also on the spectral basis in which the decomposition (\ref{normf}) has been written, reflecting the continuous freedom of these square roots again.

Ideally, we must not think in terms of numbers and vectors' components or matrices of numbers. The proper objects are the geometric and abstract algebraic entities. The matrix $g^{-1}f$ stands for a linear operator in the tangent space. Looking for its non-primary square roots, we introduce the decomposition (\ref{sei}) of a vector subspace (\ref{bei}), and if we have decided to work in coordinates, their changes must also be reflected in it. Hence, there is no violation of general covariance associated with non-primary square roots in massive gravity, though there are different problems with them.

\section{Some algebra and formal variations}

For any matrix $\mathcal M$, its first polynomial $e_1({\mathcal M})$ is, of course, nothing but its trace. So, this one is always easy to calculate. For all the other ones, $k\geqslant 2$, it is very convenient to use the Newton's recursive identity
\begin{equation}
\label{Newton}
e_k ({\mathcal M}) = -\frac{1}{k} \sum\limits_{i=1}^k (-1)^i e_1 ({\mathcal M}^i)\cdot e_{k-i}({\mathcal M}).
\end{equation}
Proving it is a simple exercise in combinatorics. I would only say that it is almost obvious. Indeed, when one wants to calculate $e_k$ by using this formula (\ref{Newton}), the first term takes $e_{k-1}$, the sum of all ordered products of $k-1$ eigenvalues, and multiplies it by the sum of all eigenvalues. It is not difficult to see that, after the proper ordering, we get $k$ times every term of $e_k$ plus some terms with one of the eigenvalues squared. The second term in the formula (\ref{Newton}) removes the unwanted squares, at the price of producing cubes if $k>2$, which are taken care of by the third term (\ref{Newton}), and so on.

In particular, the formula (\ref{Newton}) shows us that
$$e_2({\mathcal M}) = \frac12 \left(e_1^2 ({\mathcal M}) - e_1 ({\mathcal M}^2)\right).$$
One can immediately recognise that the standard Fierz-Pauli mass term around Minkowski spacetime is proportional to the second elementary symmetric polynomial of the matrix $\delta\mathcal  M = \eta^{-1} \delta g $.

Another way of calculating the polynomials $e_i$ is via the characteristic polynomial of the matrix. This is the determinant of a matrix with eigenvalues $\lambda_i - \lambda$, hence
\begin{equation}
\label{char}
\mathrm{det} ({{\mathcal M} - \lambda {\mathbb I}}) = \sum\limits_{i=1}^n (-\lambda)^{n-i} e_i(\mathcal M)
\end{equation}
with $\mathbb I$ denoting the unit matrix. From this formula (\ref{char}), we see that, if non-real eigenvalues come in complex-conjugate pairs only, then all $e_i$ are real-valued.  Of course, it also follows from a simple construction of a matrix with such spectrum over the (not algebraically closed) field of real numbers.

One more aspect worth mentioning is that the interaction term in the action (\ref{genmas}) is ($g \leftrightarrow f$)-symmetric. As long as we have chosen a particular square root $\sqrt{g^{-1}f}$, one can fix the same type of square root for $\sqrt{f^{-1}g}$ by simply defining it as 
$$\sqrt{f^{-1}g}= \left(\sqrt{g^{-1}f}\right)^{-1}.$$
Then, $e_i\left(\sqrt{f^{-1}g}\right)$ are the same polynomials as $e_i\left(\sqrt{g^{-1}f}\right)$, just in terms of $\frac{1}{\lambda_i}$ instead of $\lambda_i$. Therefore,
\begin{equation}
\label{invpol}
e_i\left(\sqrt{f^{-1}g}\right) = e_n\left(\sqrt{f^{-1}g}\right) \cdot e_{n-i}\left(\sqrt{g^{-1}f}\right),
\end{equation}
or recalling that $ e_n\left(\sqrt{f^{-1}g}\right) = \mathrm{det} \left(\sqrt{f^{-1}g}\right) =\pm \frac{\sqrt{\mathrm{-det} (g)}}{\sqrt{\mathrm{-det} (f)}}$, we find
\begin{equation}
\label{invact}
\sqrt{\mathrm{-det} (f)} \cdot e_i\left(\sqrt{f^{-1}g}\right) =\pm\sqrt{\mathrm{-det} (g)} \cdot e_{n-i}\left(\sqrt{g^{-1}f}\right),
\end{equation}
in the action principle (\ref{genmas}), with the sign depending on the sign of $\mathrm{det}\left(\sqrt{f^{-1}g}\right)$. Note that $\beta_0$ corresponds to a physical cosmological constant, while $\beta_n$ -- to the same for the metric $f$ if that is also dynamical.

One can do the variations as $\delta {\mathcal M}^2 = \delta {\mathcal M}\cdot {\mathcal M} + {\mathcal M}\cdot \delta {\mathcal M}$. Assuming 
$${\mathcal M} =\sqrt{ f^{-1}g},$$
since the trace does not depend on the order of matrices in a product, we get for a variation of $g$ with fixed $f$
$$\delta e_1 ({\mathcal M}) = \frac12 e_1({\mathcal M}^{-1}{\delta\mathcal M^2}) = \frac12 e_1 ({\mathcal M} {\mathcal M}^{-2} f^{-1} \delta g) =  \frac12 e_1 ({\mathcal M} g^{-1} \delta g),$$
so that it contributes a symmetrised $\frac12 \beta_{n-1} \sqrt{-\mathrm{det}f} \cdot {\mathcal M}g^{-1}$ term to the $\delta g$-equations. All other variations can be obtained by using the Newton's recurrence relation (\ref{Newton}) and a simple observation that
$$\delta e_1 ({\mathcal M}^k) = \frac12 e_1({\mathcal M}^{-k}{\delta\mathcal M^{2k}}) = \frac{k}{2} e_1 \left({\mathcal M}^{-k} {\mathcal M}^{2(k-1)} \delta {\mathcal M}^2\right) = \frac{k}{2} e_1 ({\mathcal M}^k g^{-1} \delta g).$$
This is enough for formally deriving the well-known field equations of massive gravity, and I will not go into any more detail on that.

In the derivations above, we have implicitly assumed that the spectral invariants of $\mathcal M$ can be found as smooth functions of the matrix $\mathcal M^2$. This is precisely what is not true with non-primary square roots. One can note \cite{nonanal} that the linearised relation
$$\delta {\mathcal M}^2 = \delta {\mathcal M}\cdot {\mathcal M} + {\mathcal M}\cdot \delta {\mathcal M}$$
is a particular case of the Sylvester equation for $\delta {\mathcal M}$, that is $\delta {\mathcal M}\cdot A + B\cdot \delta {\mathcal M} =C$ with $A=B$. There is a theorem that such equation has got a unique solution if and only if $\sigma(A)\cap \sigma(-B)=\emptyset$. This theorem is a relatively simple statement, and in our case the absence of uniqueness is almost obvious: any matrix of the form $\left(\begin{matrix}
0 & a \\
a & 0
\end{matrix}\right)$ does  anticommute with $\left(\begin{matrix}
\sqrt{\lambda} & 0 \\
0 & -\sqrt{\lambda}
\end{matrix}\right)$. It can be taken as a reminder of the huge non-uniqueness in non-primary square roots. We will see below that it leads to non-analyticity at the level of the elementary symmetric polynomials, even though they are not influenced by similarity transformations.

\section{Impossible perturbations}

The real issue with non-primary square roots is that the perturbations around them are not well-defined. Indeed, when all the eigenvalues of a matrix $\mathcal M$ are different from each other, in the sense that every Jordan block has got its own $\lambda$, not equal to any other one, there is a discrete freedom of choosing a square root $\sqrt{\mathcal M}$. This is the usual algebraic freedom of choosing the sign of $\pm\sqrt{\lambda}$. At the same time, once one gets several identical $\lambda$-s, the freedom becomes larger, with the non-primary choice of a whole algebraic variety.

As the simplest possible example, consider the $2\times 2$ unit matrix,
$${\mathbb I} \equiv \left(\begin{matrix}
1 & 0 \\
0 & 1
\end{matrix}\right).$$
Having got two equal eigenvalues, it admits two square roots of discrete freedom ($\pm\mathbb I$) when the $\sqrt{\lambda}$-s are also equal to each other, and an algebraic variety of arbitrary similarity transformations of $\left(\begin{matrix}
-1 & 0 \\
0 & 1
\end{matrix}\right)$ matrix with unequal $\sqrt{\lambda}$-s:
\begin{equation}
\label{sqr1}
\sqrt{\mathbb I} \quad = \quad \pm \left(\begin{matrix}
1 & 0 \\
0 & 1
\end{matrix}\right) \quad \mathrm{or} \quad \pm \left(\begin{matrix}
\sqrt{1-ab} & a \\
b & -\sqrt{1-ab}
\end{matrix}\right),
\end{equation}
with the former being primary and the latter -- non-primary. Since any perturbation that separates the two eigenvalues of $\mathbb I$ leaves us with only four possible square roots, it is impossible to freely perturb the matrix ${\mathcal M}=\mathbb I$ when considering the second option of its square root (\ref{sqr1}) as a continuous (in the standard topology of ${\mathbb R}^4$) function of $\mathcal M$.

For example, had we had a diagonal perturbation of $\mathcal M$, the only possible square-root matrices would be
$${\mathcal M}={\mathbb I} +\delta{\mathcal M}=\left(\begin{matrix}
1+\epsilon & 0 \\
0 & 1+\delta
\end{matrix}\right) \qquad \implies \qquad \sqrt{\mathcal M} = \left(\begin{matrix}
\pm \sqrt{1+\epsilon} & 0 \\
0 & \pm \sqrt{1+\delta}
\end{matrix}\right)$$
Therefore, if ${\mathcal M}=\eta^{-1}g$, diagonal metric perturbations around $g=\eta$ are impossible for the non-primary background square-root (\ref{sqr1}) unless $a=b=0$. In other words, around a non-primary square root, one can only consider such perturbations whose eigenvectors respect the choice of eigenvectors in the background root.

What we see is that non-primary square-root matrices can not be treated as smooth functions of the initial matrix. Let me repeat that this is easy to understand. When we take a square root with all different eigenvalues, there is no new freedom, compared to the usual square roots of complex numbers. However, when some eigenvalues coincide, there is a continuous freedom of choosing the eigenspaces of $\sqrt{\lambda}$ and $-\sqrt{\lambda}$ inside the initial eigenspace of $\lambda$, while almost every perturbation removes this freedom.

Another feature of non-primary square roots is that perturbations around them do easily lead us to complex matrices with complex-valued Lagrangian densities and equations. For example, let us take a purely off-diagonal perturbation of the 2D metric in ${\mathcal M}=\eta^{-1}g$,
$${\mathcal M} ={\mathbb I} +\delta{\mathcal M}= \left(\begin{matrix}
1 & \epsilon \\
-\epsilon & 1\end{matrix}\right).$$
The non-primary square roots (\ref{sqr1}) continue to purely imaginary matrices. Indeed, one can easily find that
$$\sqrt{\mathcal M} = \left(\begin{matrix}
 A & \frac{\epsilon}{2A} \\
-\frac{\epsilon}{2A} & A
\end{matrix}\right), \qquad A=\pm \sqrt{\frac12 \left(1\pm \sqrt{1+\epsilon^2}\right)}$$
with two real and two imaginary branches. For the imaginary ones, $e_1(\sqrt{\mathcal M})\approx \pm i\epsilon$ and
$$\pm\frac{1}{\sqrt{2}} \left(\begin{matrix}
 \sqrt{1- \sqrt{1+\epsilon^2}} & {\epsilon}/{ \sqrt{1- \sqrt{1+\epsilon^2}}} \\
-{\epsilon}/{ \sqrt{1- \sqrt{1+\epsilon^2}}} &  \sqrt{1- \sqrt{1+\epsilon^2}}
\end{matrix}\right) \quad {\mathop {\longrightarrow}\limits_{\epsilon\to 0}}\quad \pm \left(\begin{matrix}
0 & i \\
-i & 0
\end{matrix}\right).$$
Therefore, this case corresponds to non-primary square roots (\ref{sqr1}) of $a=-b=\pm i$.

One can check that the origin of this complex trouble is in choosing the square roots of two complex-conjugate eigenvalues $\lambda = 1\pm i\epsilon$ that are no longer complex-conjugate to each other. This simple fact is enough for excluding the non-principal square roots for most of the physical applications of massive gravity. We are often interested in backgrounds of high symmetry, and quite probably with some equal eigenvalues. If two equal eigenvalues were separated in their square roots, it might happen that some perturbations turn them into a complex-conjugate pair and produce a complex-valued square-root matrix out of that.

\section{Formulation without the square-root matrices}

If to forget about the unwanted complex numbers, one would naturally want to explore the non-primary square roots. Even though the initial construction \cite{dRGT} was obtained  by summing a power series around $\sqrt{\mathbb I}={\mathbb I}$, the ghost-freedom proof \cite{HR2} does not assume any background. Moreover, one can attack the issue \cite{memas} by introducing the $e_i(\Phi)$-terms and a Lagrange multiplier term setting $\Phi^2=g^{-1}f$.

We have proposed a method of working with the elementary symmetric polynomials alone, without explicit square-root matrices \cite{we1,we3}. Note that the polynomials $e_i$ do not depend on any similarity transformation. At the same time, the square-root matrix $\sqrt{g^{-1}f}$ does not enter anywhere in the action functional (\ref{genmas}), except of the $e_i$ interaction terms. Therefore, the continuous freedom of the non-primary square roots, together with the issues of their covariance, are absolutely irrelevant to the physical content of the model.

One can find the possible eigenvalues of $\sqrt{\mathcal M}$ even without looking for the matrix itself. At the level of the elementary symmetric polynomials, it is a simple algebraic task \cite{we1}. Indeed, starting from the characteristic polynomial (\ref{char}), we observe that
$$\mathrm{det}\left( {\mathcal M}^2 - \lambda^2 {\mathbb I} \right)= \mathrm{det}\left( {\mathcal M} - \lambda {\mathbb I}\right) \cdot \mathrm{det}\left( {\mathcal M} + \lambda {\mathbb I}\right)$$
and, equating the coefficients in front of the even powers of $\lambda$, we get
\begin{equation}
\label{quadrrel}
e_k({\mathcal M}^2) = \sum_{i,j:\quad i+j=2k} (-1)^{\frac{i-j}{2}} \cdot e_i({\mathcal M}) \cdot e_j({\mathcal M}).
\end{equation}
Recalling that $e_i=0$ for $i<0$ and for $i>n$, we see that (\ref{quadrrel}) is a finite system of equations consistent with the facts that $e_0=1$ and  $e_n({\mathcal M}^2) = e_n^2({\mathcal M})$.

Therefore, one can formulate the massive gravity theory (\ref{genmas}) as an action principle with a number of implicit functions ${\mathfrak e}_k (g,f)$ in it \cite{we1}:
\begin{equation}
\label{polgenmas}
S=\int  \left(R(g) + \sum\limits_{k=0}^n \beta_k {\mathfrak e}_k\right)\sqrt{\mathrm{-det} (g)}  d^n x; \qquad\qquad \sum_{i,j:\quad i+j=2k} (-1)^{\frac{i-j}{2}}  {\mathfrak e}_i {\mathfrak e}_j \equiv e_k (g^{-1}f).
\end{equation}
with the additional condition of ${\mathfrak e}_0=1$. In particular, in 4D the quadratic equations (\ref{polgenmas}) are
\begin{equation}
\label{4Dquad}
 {\mathfrak e}_1^2 - 2 {\mathfrak e}_2 = e_1, \qquad {\mathfrak e}_2^2 - 2{\mathfrak e}_1 {\mathfrak e}_3 + 2{\mathfrak e}_4= e_2, \qquad {\mathfrak e}_3^2 - 2 {\mathfrak e}_2 {\mathfrak e}_4 = e_3, \qquad {\mathfrak e}_4^2 = e_4.
\end{equation}

As an example, one can take the $4\times 4$ unit matrix $\mathbb I$ with $e_i=(4,6,4,1)$ and find that all solutions of equations (\ref{4Dquad}) for ${\mathfrak e}_i$ are $(4,6,4,1)$ for the $\sqrt{\mathbb I}=\mathbb I$ principal root, $(2,0,-2,-1)$ for a non-primary with one sign reversed, $(0,-2,0,1)$ with two signs reversed, $(-2,0,2,-1)$ with three signs reversed, $(-4,6,-4,1)$ for $\sqrt{\mathbb I}=-\mathbb I$. If a solution for some $e_i$ is known, one can study perturbations around it \cite{we1}.

\section{The lack of analyticity}

The non-analyticity for perturbations around non-primary square roots was noticed in \cite{nonanal} and further investigated in \cite{meunr}. Basically, the matrix of $\frac{\partial e_i}{\partial {\mathfrak e}_j}$ is degenerate on the non-primary square roots, and therefore, finding $\delta{\mathfrak e}_i$ in terms of $\delta e_i$ cannot go smoothly.

Just for illustration, consider the case of 2D. The algebraic equations (\ref{4Dquad}) take a very simple form:
$$ {\mathfrak e}_1^2 - 2 {\mathfrak e}_2 = e_1, \qquad {\mathfrak e}_2^2 = e_2.$$
If we take $\sqrt{\mathbb I}= \left(\begin{matrix}
-1 & 0 \\
0 & 1
\end{matrix}\right)$, it means $e_1=2$, $e_2=1$ and ${\mathfrak e}_1=0$, ${\mathfrak e}_2=-1$. Perturbing this solution, we get
$${\mathfrak e}_2= - \sqrt{e_2}, \qquad  {\mathfrak e}_1=\pm \sqrt{e_1 - 2\sqrt{e_2}}.$$
The meaning of the two possible signs in ${\mathfrak e}_1$ is obvious: it depends on whether the bigger or the smaller perturbed eigenvalue goes with the minus sign. This information is not in the background matrix.

Denoting the perturbed matrix as ${\mathbb I} + {\mathcal H}$, one can calculate in components
$${\mathfrak e}_1 = \pm\frac12 \sqrt{2  {\mathcal H}^{\mu}_{\nu}  {\mathcal H}^{\nu}_{\mu} - \left( {\mathcal H}^{\mu}_{\mu}\right)^2} + {\mathcal O}\left( {\mathcal H}^2\right),$$
the non-analytic behaviour at ${\mathcal H}\to 0$. Taking into account that ${\mathcal H}^0_1 = - {\mathcal H}^1_0$ if ${\mathcal H}=\eta^{-1}\delta g$, the square-root trace is real for diagonal metric perturbations and imaginary for the strictly off-diagonal ones as we've seen above, with non-analytic behaviour when crossing the locus of $2  {\mathcal H}^{\mu}_{\nu}  {\mathcal H}^{\nu}_{\mu} - \left( {\mathcal H}^{\mu}_{\mu}\right)^2 =0$, or more precisely when $e_1=2\sqrt{e_2}$ for the perturbed matrix, too, that happens when ${\mathcal H} \propto \mathbb I$.

We have seen the same situation in all other cases \cite{meunr}: a cubic equation for the first order perturbation $\delta{\mathfrak e}_1$ around $\sqrt{\mathbb I}= \mathrm{diag} (-1, 1, 1)$, a quartic one around $\sqrt{\mathbb I}= \mathrm{diag} (-1, 1, 1, 1)$, and a cubic equation for ${\mathfrak e}_1^2$ for perturbations around $\sqrt{\mathbb I}= \mathrm{diag} (-1, -1, 1, 1)$. In all the cases, other quantities can be easily found, once the first-order perturbation of ${\mathfrak e}_1$ is known and chosen, and the order of its equation can be understood from the need of three, four, or six different solutions depending on the choice of eigenvalue(s) with reversed sign.

\section{Minkowski space with Lorentz violation}

Every time we use a non-primary square root, there is a danger of getting complexified perturbed square roots. There is no hope of being protected from that in physically interesting situations. Indeed, whatever Ans{\"a}tze of proportional solutions we might take, including the weak massive gravity limit with Minkowski fiducial metric, we have $g^{-1}f = c^2\cdot\mathbb I$. A  non-primary square root, in any sense respecting rotational invariance, would take then the form of $\mathrm{diag}\left(-c,+c,+c,+c\right)$, or analogously in any other dimension. Mixing spatial dimensions with the temporal one, we can construct a perturbation leading to complex numbers as above.

One can, of course, choose a matrix of $f^{-1}g = \mathrm{diag}(c^2, 1, 1, 1)$ as in the paper \cite{nonanal}. Then, as long as $c^2\neq 1$, the perturbation theory is analytic, even though quite complicated in calculations. However, the limit of $c\to 1$ is singular. And there is no reason to believe that such limits are never to be approached, especially in a bimetric theory.

For illustration purposes, let me consider perturbations around 4D unit matrix with $\sqrt{\mathbb I}=\mathrm{diag}(-1, 1, 1, 1)$. Let us assume that the reversed-sign eingenvector is the time-like one, and simply denote the eigenvalues of $\eta^{-1}\delta g$ by $\delta_{\mu}$. Then the eigenvalues of $\sqrt{\eta^{-1}g}$ must be $\lambda_0 =-\sqrt{1+\delta_0}$ and $\lambda_i =\sqrt{1+\delta_i}$. I will use the approximation of $\sqrt{1+\delta} =1+ \frac12 \delta - \frac18 \delta^2 + {\mathcal O}(\delta^3)$ and introduce a notation of ${\tilde\delta}\equiv\delta-\frac14 \delta^2$.

By straightforwardly calculating the elementary symmetric polynomials (\ref{defe}) of these $\lambda$-s, we get
\begin{multline}
\label{bigpert}
\beta_3 {\mathfrak e}_1 + \beta_2{\mathfrak e}_2 + \beta_1 {\mathfrak e}_3 + \beta_0 {\mathfrak e}_4 = 2(\beta_3-\beta_1)-\beta_0 \\
- {\tilde\delta}_0 \cdot \frac12 \left(\beta_3+3\beta_2+3\beta_1 +\beta_0\right)
+({\tilde\delta}_1 + {\tilde\delta}_2 + {\tilde\delta}_3)\cdot\frac12\left(\beta_3+\beta_2-\beta_1- \beta_0\right) \\
- \delta_0(\delta_1 + \delta_2 + \delta_3)\cdot \frac14 \left(\beta_2+2\beta_1+\beta_0\right)
+(\delta_1 \delta_2 + \delta_1 \delta_3 + \delta_2 \delta_3 ) \cdot \frac14 \left(\beta_2 - \beta_0 \right) + {\mathcal O}(\delta^3)
\end{multline}
for minus (since the determinant in (\ref{invpol}) is negative) the interaction density (\ref{invact}).

For a Minkowski space to be a solution, we need to make the linear terms (\ref{bigpert}) vanish:
\begin{equation}
\label{Minksol}
\beta_3+3\beta_2+3\beta_1 +\beta_0 =0, \qquad \beta_3+\beta_2-\beta_1- \beta_0 =0.
\end{equation}
The first equality is the standard condition for the Minkowski space to be a solution, with the principal square root. The second equality is a new one, and it would also appear \cite{nonanal} for the $g=\eta $ solution with a fiducial metric $f=\mathrm{diag}\left(-A, +1, +1, +1\right)$ of arbitrary $A\neq 1$. I prefer to write the conditions (\ref{Minksol}) in a more symmetric way:
\begin{equation}
\label{Minksym}
\beta_0+2\beta_1 +\beta_2 =0, \qquad \beta_1+2\beta_2+\beta_3 =0.
\end{equation}

Taking the condition (\ref{Minksym}) into account, we find the interactions' contribution (\ref{bigpert}) to the quadratic Lagrangian around the non-primary square root:
\begin{equation}
\label{Lagpert}
\delta {\mathcal L}_{\mathrm{int}} =
-\frac{\beta_1 + \beta_2}{2} \cdot (\delta_1 \delta_2 + \delta_1 \delta_3 + \delta_2 \delta_3 ) + {\mathcal O}(\delta^3).
\end{equation}
Previously \cite{meunr}, we obtained this result (\ref{Lagpert}) by a very cumbersome calculation of perturbing the equations (\ref{4Dquad}) and checking that the solutions of the quartic equation for the first-order approximation to $\delta{\mathfrak e}_1$, according to Vieta's formulae, are nothing but $\frac12 (-\delta_0 + \delta_1 + \delta_2 + \delta_3)$ and the three others of the same type.

This result (\ref{Lagpert}) explains the complicated (and non-analytic) nature of perturbative calculations \cite{meunr}. It all must be as complicated as to precisely remove one of the eigenvalues from the simple expression of $e_2(\mathcal H)$. Note also that we are back to the issue of complex-valued Lagrangians. Indeed, if we take $\delta g_{01}=\epsilon_1$, $\delta g_{22}=\epsilon_2$, and all other components vanishing, the perturbation eigenvalues are $\pm i \epsilon_1$, $\epsilon_2$, $0$. Having removed one of the first two of them (assuming that it represents the reversed-sign eigenvalue of $\sqrt{\eta^{-1}g}$), we get the quadratic interaction Lagrangian (\ref{Lagpert}) taking a value propotional to $i \epsilon_1 \epsilon_2$.

Actually, it seems that having taken a non-primary square root of $\mathbb I$ and separating the vector space  into two invariant subspaces ${\mathbb V}_{\lambda\approx +1} \bigoplus {\mathbb V}_{\lambda\approx -1}$ for the operator $\sqrt{{\mathbb I}+{\mathcal H}}$, the quadratic interaction Lagrangian takes the form of two pieces \cite{meunr}, one proportional to $e_2 (\mathcal H)$ in ${\mathbb V}_{\lambda\approx +1}$ and another proportional to $e_2 (\mathcal H)$ in ${\mathbb V}_{\lambda\approx -1}$. For sure, each one of them can be non-trivial only if the corresponding subspace is of dimension two or more.

\section{Conclusions}

There are many interesting issues about the square-root matrices of massive and bimetric gravity. In particular, a topic I have not touched upon here at all is a question of when a real-valued matrix admits at least one real-valued square root \cite{MiKo}. If we were all right about an isolated solution with no perturbations around, the non-primary square roots could help us. Indeed, if there are two identical negative eigenvalues in the matrix, one can take a complex-conjugate pair of their square roots and construct a real square-root matrix, if there are no other obstructions to that.

If we want to build a perturbation theory, as we should, the non-primary square roots are not good, due to the bad analytic properties described above; and even going to the matrix-less formulation in terms of elementary symmetric polynomials \cite{we1} does not help much with that. In this sense, I do absolutely agree with the preference for the principal square roots \cite{MiKo}. However, I hope that I have shown in this paper that the problems of other square roots do not have anything to do with the issues of general covariance per se.

\end{document}